\documentclass[useAMS,usenatbib]{mn2e}
\usepackage{graphicx}
\usepackage{natbib}
\usepackage{lscape}
\usepackage{longtable}
\usepackage{subfig}
\usepackage{amssymb,amsmath}
\usepackage{times}
\usepackage[T1]{fontenc}
\usepackage{hyperref}
\usepackage{amsmath}
\usepackage{breakurl}
\usepackage{times}
\usepackage{morefloats}
\usepackage{bm}
\usepackage{flafter}

\newcommand{\mbh}{$M_{\textrm{bh}}$}
\newcommand{\mhalo}{$M_{\textrm{halo}}$}
\newcommand{\mstar}{$M_{\textrm{stellar}}$}
\newcommand{\fedd}{$\lambda_{\textrm{Edd}}$}
\newcommand{\lcut}{$L_{\textrm{cut}}$}
\newcommand{\tobsc}{$\tau_{\textrm{obsc}}$}
\newcommand{\meanfedd}{$\langle \lambda_{\textrm{Edd}} \rangle$}
\newcommand{\mswitch}{$\Delta M_{\textrm{switch}}$}
\newcommand{\fobsc}{$f_{\textrm{obsc}}$}

\title[Obscured quasar haloes and evolution]{A unifying evolutionary framework for infrared-selected obscured and unobscured quasar host haloes}
\author[DiPompeo et al.]{M.A.\ DiPompeo$^{1}$, R.C. Hickox$^{1}$, A.D. Myers$^{2}$, J.E. Geach$^{3}$ \\
$^1$Department of Physics and Astronomy, Dartmouth College, 6127 Wilder Laboratory, Hanover, NH 03755, USA \\
$^2$Department of Physics and Astronomy 3905, University of Wyoming, 1000 E. University, Laramie, WY 82071, USA \\
$^3$Centre for Astrophysics Research, Science \& Technology Research Institute, University of Hertfordshire, Hatfield, AL10 9AB, UK}

\begin{document}

\date{Accepted ?; ?; in original form 2016 May 1}

\pagerange{\pageref{firstpage}--\pageref{lastpage}} \pubyear{2016}

\maketitle

\label{firstpage}

\begin{abstract}
Recent measurements of the dark matter halo masses of infrared-selected obscured quasars are in tension --- some indicate that obscured quasars have higher halo mass compared to their unobscured counterparts, while others find no difference.  The former result is inconsistent with the simplest models of quasar unification that rely solely on viewing angle, while the latter may support such models.  Here, using empirical relationships between dark matter halo and supermassive black hole masses, we provide a simple evolutionary picture that naturally explains these findings and is motivated by more sophisticated merger-driven quasar fueling models.  The model tracks the growth rate of haloes, with the black hole growing in spurts of quasar activity in order to ``catch-up'' with the \mbh-\mstar-\mhalo\ relationship.  The first part of the quasar phase is obscured and is followed by an unobscured phase.  Depending on the luminosity limit of the sample, driven by observational selection effects, a difference in halo masses may or may not be significant.  For high luminosity samples, the difference can be large (a few to 10 times higher masses in obscured quasars), while for lower luminosity samples the halo mass difference is very small, much smaller than current observational constraints.  Such a simple model provides a qualitative explanation for the higher mass haloes of obscured quasars, as well as rough quantitative agreement with seemingly disparate results. 
\end{abstract}


\section{INTRODUCTION}
The advent of large infrared (IR) astronomical surveys, beginning with \textit{Spitzer} \citep{2004ApJS..154....1W} and more recently with the \textit{Wide-field Infrared Survey Explorer} \citep[\textit{WISE};][]{2010AJ....140.1868W}, has begun to draw back the curtain on the obscured phase of black hole growth in bursts of quasar\footnote{Quasars are the highly-luminous fraction of the more broad classification of active galactic nuclei (AGN) --- we will use the terms interchangeably here, without regard to specific luminosity divisions.} activity \citep[e.g.][]{2004ApJS..154..166L, 2004ApJ...616..123T, Stern:2005p2563, 2007ApJ...671.1365H, 2009ApJ...696..891H, 2011ApJ...731..117H, 2012ApJ...753...30S, 2014ApJ...795..124H, 2014MNRAS.442.3443D, 2014ApJ...782....9H, 2015ApJ...802..102L, 2015MNRAS.446.3492D, 2016MNRAS.456..924D} that has historically been dominated by optically detected unobscured systems \citep[e.g.][]{2004MNRAS.349.1397C, 2005MNRAS.356..415C, Richards:2006p3932, 2006AJ....131.2766R, 2007ApJ...658...85M, 2009MNRAS.399.1755C, 2009ApJ...697.1634R, 2009ApJ...697.1656S, 2009MNRAS.396..423B, 2015MNRAS.453.2779E}.  This has resulted in tension between the prevailing paradigms that attempt to explain the physical origin of observational quasar subclasses.  On the one hand, orientation effects due to the non-spherically symmetric geometry of quasars can strongly impact observations \citep[e.g.][]{2013MNRAS.429..135R, 2013MNRAS.435.3251R, 2014ApJ...787...73D, 2014MNRAS.438.3263R, 2015MNRAS.454.3864B, 2016arXiv160206954S} and in the presence of an axis-symmetric dust distribution (the so-called ``dusty-torus''), orientation alone has the potential to unify obscured and unobscured quasars, as it largely has for Seyfert galaxies \citep[e.g.][]{1993ARA&A..31..473A, 2012MNRAS.420.2756S, 2015ApJ...803...57I, 2015A&A...581L...8G, 2015P&SS..116...97M, 2015ARA&A..53..365N, Siebenmorgen:2015wc, 2015ApJ...812...99L, 2016MNRAS.457..745W, 2016MNRAS.456.2861O}.  On the other, attempts to explain the driving of gas into nuclear regions in order to fuel the supermassive black hole (SMBH) have resulted in models that link quasar activity to major galaxy mergers and a general evolutionary sequence \citep[e.g.][]{1988ApJ...325...74S, 2000MNRAS.311..576K, 2003ApJ...596L..27K, 2003ApJ...595..614W, 2004ApJ...600..580G, 2006MNRAS.365...11C, Hopkins:2006p654, 2009MNRAS.398...53B, 2009ApJ...698.1550H, 2009ApJ...690...20S, 2013ApJ...762...70C, 2013MNRAS.428..421S}. 

Some evolutionary models predict that obscured and unobscured quasars can have different parent dark matter halo masses, and such a difference is in direct contradiction to pure unification by orientation.  With large statistical samples of obscured quasars now available, attention has turned to probing this question via quasar clustering and CMB lensing cross-correlations, which can be used to infer typical dark matter halo masses.  Clustering measurements of IR-selected samples have resulted in contradictory results.  \citet{2011ApJ...731..117H} and later \citet{2014MNRAS.442.3443D} and \citet[][hereafter D16a]{2016MNRAS.456..924D} used angular cross- and auto-correlations, respectively, to show that indeed obscured quasars reside in higher mass haloes than unobscured quasars selected in a similar way. \citet{2014ApJ...789...44D} also found that obscured quasars are found in higher mass haloes, though at a much more extreme level, seemingly due to insufficient masking of the \emph{WISE} data \citep{2014MNRAS.442.3443D}. However, \citet{2016ApJ...821...55M} used a smaller sample with spectroscopic redshifts to make a projected correlation measurement, and found no difference between obscured and unobscured haloes.  It should be noted that the luminosities probed by the \citet{2016ApJ...821...55M} sample extend to lower values than in D16a, a point which we will return to later in this work.

Following up on these clustering measurements, \citet{2015MNRAS.446.3492D} and D16a used a cross-correlation with \textit{Planck} CMB lensing maps \citep{2014A&A...571A..17P, PlanckCollaboration:2015tp} to measure the halo masses of the same obscured and unobscured quasar samples used for clustering measurements.  This follow up with an independent measurement technique provided remarkably consistent results, with obscured quasars residing in haloes 2-3 times larger than than those of unobscured quasars.  

\citet[][hereafter D16b]{2016MNRAS.460..175D} argued that obscured samples selected in the mid-IR are really a mix of torus-obscured sources intrinsically identical to unobscured sources, since it is fairly well established from models and observations that a structure like the dusty torus exists \citep[e.g.][]{2006A&A...452..459H, 2008ApJ...685..147N, 2008ApJ...685..160N, Deo:2011p1917, 2011MNRAS.414.1082M, 2012MNRAS.420.2756S, 2014ApJ...794..111B, 2016MNRAS.455.3968H, 2016MNRAS.456L..94M, 2015MNRAS.451.2991G}, and those obscured by some other dust distribution \citep[``non-torus obscured'', or NTO, quasars, e.g.][]{2012ApJ...755....5G}.  It is this latter subset that are of interest in an evolutionary framework, while the former are unified by simple orientation --- i.e.\ explaining the full population requires both orientation and evolution.  Because torus-obscured objects should have similar halo masses to unobscured sources, their presence in the obscured sample actually dilutes the non-torus obscured signal, which could therefore have halo masses nearly 10 times larger than unobscured quasars (D16b).

\citet{2011ApJ...731..117H}, \citet{2014MNRAS.442.3443D, 2015MNRAS.446.3492D}, and D16a attempted to explain the larger obscured halo masses with an evolutionary picture in which black hole growth lags behind dark matter halo growth \citep[e.g.][]{2006ApJ...649..616P, 2008AJ....135.1968A, 2008ApJ...681..925W, 2010MNRAS.402.2453D, 2013ARA&A..51..511K} and the obscured phase precedes the unobscured phase \citep[e.g.][]{Hopkins:2006p654, 2009MNRAS.398...53B}, with the unobscured and obscured phases lasting roughly the same length of time (each on the order of 100 Myr, or 1\% of the Hubble time).  Because the obscured and unobscured samples have similar luminosities \citep[e.g.][]{2007ApJ...671.1365H}, they should have equally well-matched black hole masses (assuming similar Eddington ratios).  However, since the obscured phase precedes the unobscured phase, the black holes harbored by obscured quasars are under-massive relative to their haloes compared to the unobscured sources, i.e.\ they are ``catching up'' to their final black hole masses.  Therefore, obscured sources are preferentially observed in higher mass haloes. 
%\citep[][]{2013ApJ...772...26A, 2013ApJS..208...24L, 2013MNRAS.434..941M, 2013AJ....145...55Y, 2014ApJ...795..124H}.

In this work, we present a simple model to illustrate that this general evolutionary picture in the context of empirical relationships between BH masses, galaxies, and haloes, can naturally produce the observed difference between the haloes that host obscured and unobscured quasars.  In addition, we aim to illustrate that apparent disagreements in the literature may be explained by differences in selection based on luminosity.

We adopt a cosmology of $H_0 = 70.2$ km s$^{-1}$ Mpc$^{-1}$,  $\Omega_{\textrm{M}} =  \Omega_{\textrm{CDM}} +  \Omega_{\textrm{b}} = 0.229 + 0.046 = 0.275$, $\Omega_{\Lambda} = 0.725$, and $\sigma_8 = 0.82$ \citep{2011ApJS..192...18K}.

\section{The Model}
\subsection{Dark Matter Halo, Galaxy, and Black Hole Growth}
We begin by generating a sample of 5000 dark matter haloes at $z=3$ \citep[the maximum redshift at which IR colors are effective for selecting quasars, e.g.][]{2012ApJ...753...30S, 2013ApJ...772...26A}.  The masses are uniformly distributed in the logarithm of mass between $10^{11}$ M$_{\odot}/h$, more than a dex lower than the typical halo masses of quasars, and $10^{15}$ M$_{\odot}/h$, masses which are exceedingly rare.  \citet{2010MNRAS.406.2267F} traced the halo merger trees of the \textsc{millennium-II} simulation \citep{2009MNRAS.398.1150B} and fit a functional form to the halo mass growth rate $dM_{h}/dz$.  Adopting their Equation 2 for the median halo growth rate, we grow each halo from $z=3$ to $z=0$ (in steps of 0.001), storing the masses at each step.

\begin{figure}
\centering
   \includegraphics[width=8.cm]{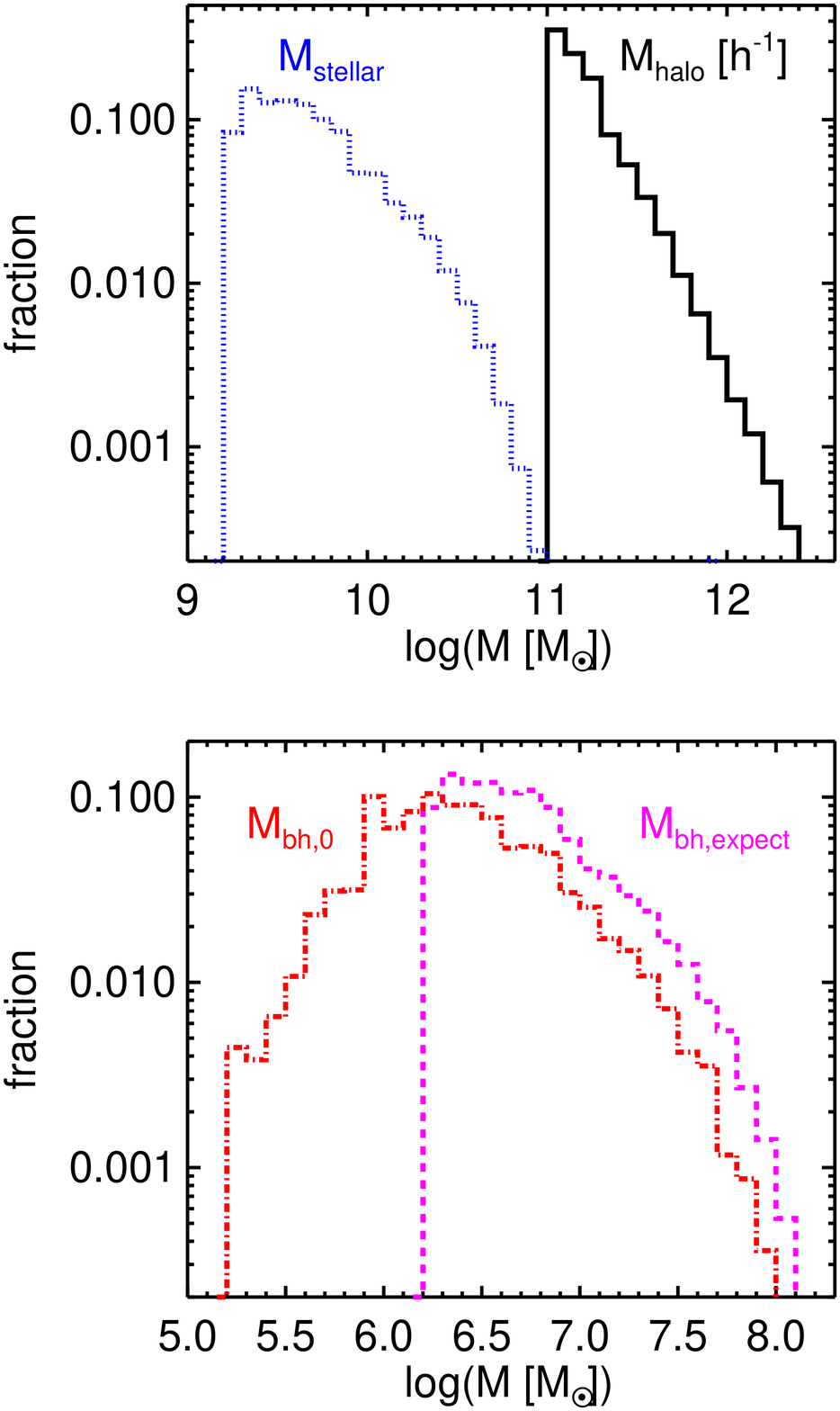}
  \caption{The normalized initial $z=3$ mass distributions, weighted by halo mass according to the $z=3$ halo mass function of \citet{2010ApJ...724..878T}.  \emph{Top:} \mstar\ (dotted blue) and \mhalo\ (solid black).  Note that we retain the $h^{-1}$ units on the halo masses for consistency with convention. \emph{Bottom:} The expected black hole masses $M_{\textrm{bh,expect}}$ (dashed magenta) are generated according to the relationships shown in Figure~\ref{fig:mass_rels}, and the actual initial BH masses $M_{\textrm{bh,0}}$ (dot-dashed red) are simply a random fraction from 0.1 to 0.9 of each expected mass.\label{fig:in_masses}
  %The normalized initial $z=3$ mass distributions, truncated at $\log(M_{\textrm{halo}}) = 11.5$ for clarity and weighted according the the $z=3$ halo mass function of \citet{2010ApJ...724..878T}.  \mstar\ (dotted blue) and $M_{\textrm{bh,expect}}$ (dashed magenta) are generated according to the relationships shown in Figure~\ref{fig:mass_rels}. The initial BH masses $M_{\textrm{bh,0}}$ (dot-dashed red) are simply a random fraction from 0.1 to 1 of each expected BH mass.\label{fig:in_masses}
  }
\end{figure}

A weight is assigned to each halo in accordance with the $z=3$ halo mass function (HMF) of \citet{2010ApJ...724..878T}.  We weight each halo as opposed to generating an initial distribution consistent with the HMF in order to remove mass-dependent shot noise that can skew the results when considering high-mass haloes, without having to generate a much larger initial sample and increasing computational time.  Once a weight is assigned, each object retains its weight despite its growing mass and the evolving HMF, as evolving the weights complicates the model without significantly altering the final results.  This is because over the range of typical halo masses of quasars, the slope of the HMF does not change significantly over the redshift range of the final mock samples ($0.5 < z < 1.5$; see Section 2.2), as well as the general rarity of the most massive halos at any redshift.  The weighted initial mass distribution of haloes is shown in Figure~\ref{fig:in_masses}.  Note that throughout this work halo masses are always presented in units of $M_{\odot}/h$, in keeping with convention, while other masses are given in units of $M_{\odot}$.

\begin{figure}
\centering
   \includegraphics[width=7cm]{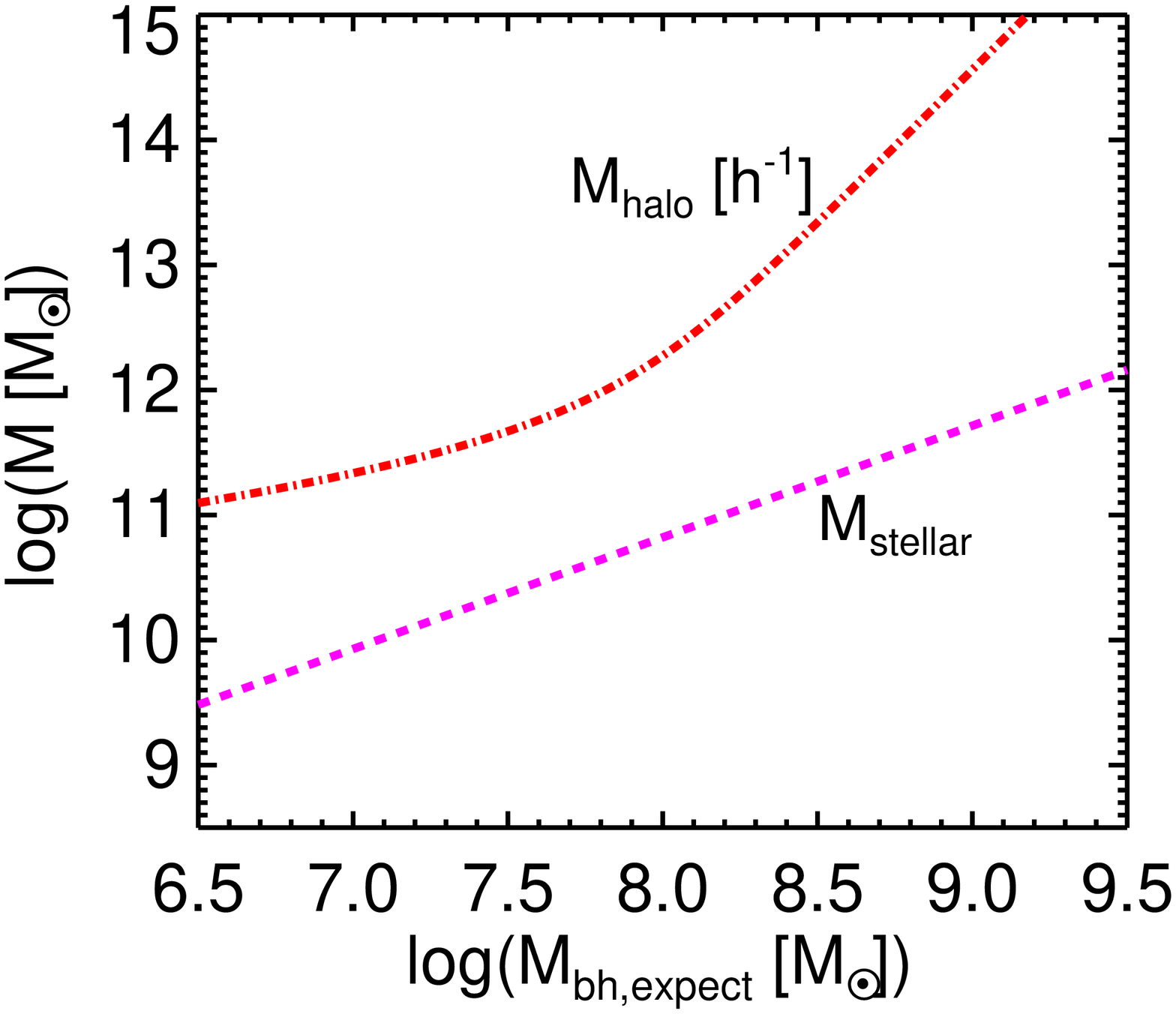}
  \caption{The relationships between the expected BH mass and the halo and stellar masses of each source.  These curves are defined by the \mhalo\ - \mstar\ relationship \citep[][]{2010MNRAS.404.1111G} and \mstar\ - \mbh\ relationpship \citep[][]{2004ApJ...604L..89H}.\label{fig:mass_rels}}
\end{figure}

Using the relationship between galaxy stellar mass and host halo mass of \citet{2010MNRAS.404.1111G}, shown here in the top panel of Figure~\ref{fig:mass_rels}, we predict the total stellar mass \mstar\ of a potential quasar host galaxy within each halo at each redshift step.  The distribution of expected stellar masses for the initial haloes is shown in Figure~\ref{fig:in_masses}.  We then predict the expected SMBH mass of each galaxy using the stellar mass - BH mass relation of \citet{2004ApJ...604L..89H}, which is shown in the bottom panel of Figure~\ref{fig:mass_rels}. The distribution of initial expected BH masses for each galaxy is shown in Figure~\ref{fig:in_masses}.  We have assumed no evolution in the \mhalo\ - \mstar\ relationship, as we are interested in $z<3$ \citep[e.g.][]{2013ApJ...762L..31B, 2013ApJ...770...57B}.  We further assumed that the \mstar\ - \mbh\ relationship does not evolve with cosmic time, which may or may not be the case \citep[e.g.][]{2010MNRAS.402.2453D} as selection effects and biases makes this relationship difficult to probe as redshift increases.  However, such an evolution, if it exists, appears to be similar in strength to the scatter in the mass relationships at a given time.  Given that we are ignoring the scatter, both for simplicity and because the intrinsic amount is difficult to establish, we also choose to assume no evolution.

Assuming that BHs grow in bursts of accretion activity, the initial $z=3$ distribution of BH masses will not be identical to the expected masses predicted from the \mhalo\ - \mstar\ - \mbh\ relationships, as not all objects have grown onto this plane \citep[under the assumption that BH growth lags behind halo growth, see e.g.][]{2006ApJ...649..616P, 2008AJ....135.1968A, 2008ApJ...681..925W, 2010MNRAS.402.2453D, 2013ARA&A..51..511K}.  To generate a starting point, we simply begin each BH at a random fraction (from 0.1 to 0.9) of its expected value (Figure~\ref{fig:in_masses}).

\begin{figure}
\centering
   \includegraphics[trim=0cm -0.2cm 0cm 0cm, clip, width=8cm]{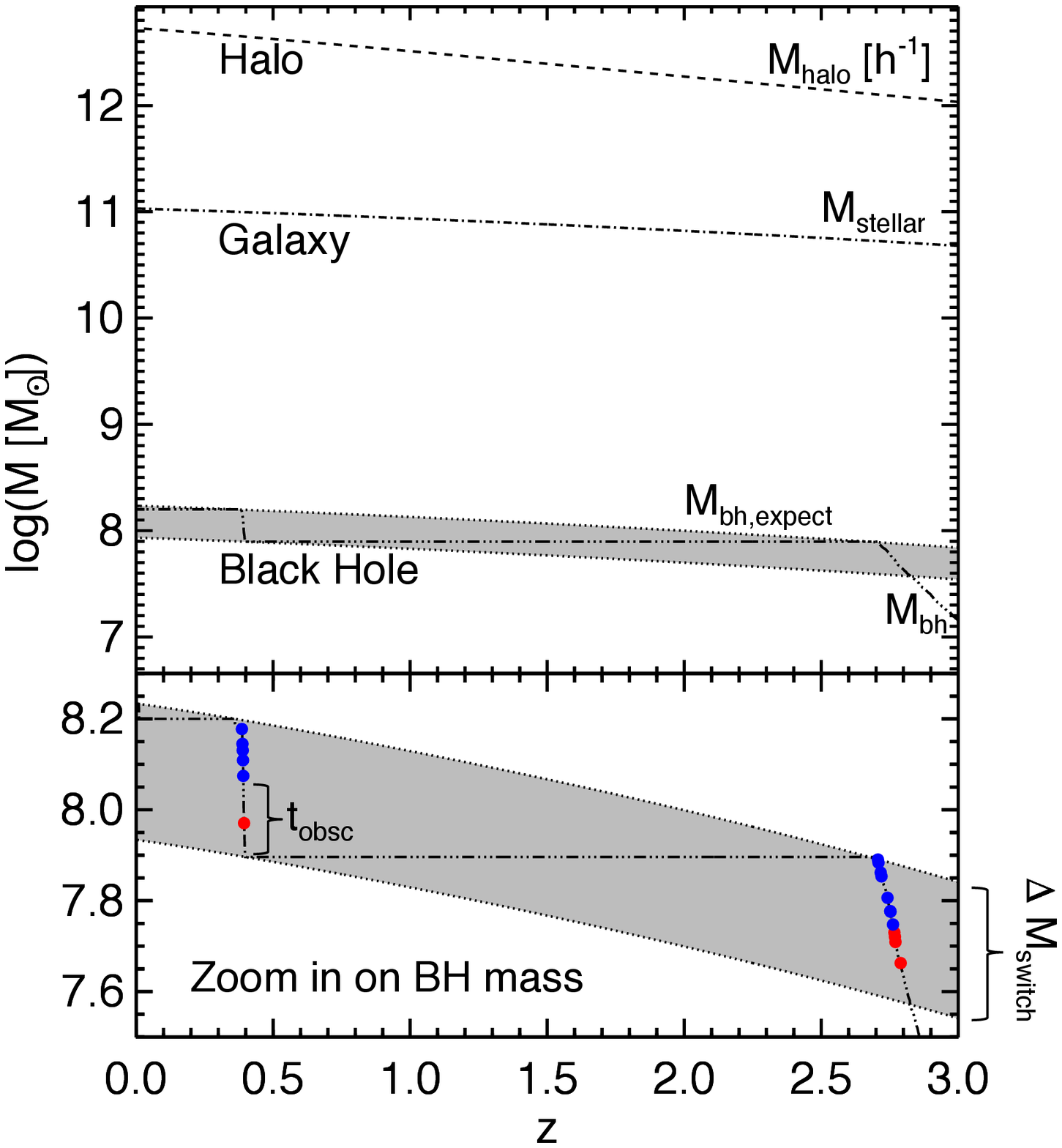}
   %\vspace{0.5cm}
  \caption{An example of the growth of an individual source.  \emph{Top:} The halo mass grows at the median rate of \citet{2010MNRAS.406.2267F}, and the stellar and expected BH masses are related to the halo mass at each step by the relationships in Figure~\ref{fig:mass_rels}.  If the mass of a BH falls outside of the grey region defined by \mswitch\ it enters a phase of growth until it reaches its expected mass.  \emph{Bottom:} Mass growth focused on just the BH.  For this example, an \fedd\ is randomly assigned from a Schechter function at each step, and combined with \mbh\ determines the luminosity.  The first part of each growth phase is assumed to be obscured and the later part unobscured.  The red and blue points indicate where in this object's growth it satisfies the luminosity cut (section 2.2) and is observed as obscured and unobscured, respectively.  Gaps in the colored points indicate ``flickering'' due to the varying \fedd.  In the single \fedd\ case, the entire growth phase would be visible down to a minimum mass, corresponding to the luminosity cut.\label{fig:growth}}
\end{figure}

In our model, the BH begins to grow if its mass falls below a specified fraction of its expected value. We define the mass ratio $M_{\textrm{bh}}/M_{\textrm{bh,expect}}$ at which an AGN phase begins as \mswitch, and allow its value to be a free parameter in the model.  When an AGN phase begins, the BH grows according to the Salpeter (or $e$-folding) time \citep{1964ApJ...140..796S}:
\begin{equation}
t_s = \frac{M_{\textrm{bh}}}{\dot{M}_{\textrm{bh}}} = 4.5 \times 10^7 \left( \frac{\epsilon}{0.1} \right) \left( \frac{L}{L_{\textrm{Edd}}} \right)^{-1} \ \textrm{yr},
\end{equation}
where $\epsilon$ is the radiative efficiency (assumed to be 0.1), $L$ is the AGN bolometric luminosity, and $L_{\textrm{Edd}} = 1.26 \times 10^{38} (M_{\textrm{bh}}/M_{\odot})$ ergs s$^{-1}$ is the Eddington luminosity.  The bolometric luminosity is determined by multiplying $L_{\textrm{Edd}}$ by the Eddington ratio \fedd\, which is treated in two ways that will be compared in Section 3.4.  The first is to simply assume an average \fedd\ and assign all sources this average value at all times, which simplifies our prescription and allows discussion of the model with one fewer complicating factor.  However, quasars exhibit a range of Eddington ratios \citep[e.g.][]{2009ApJ...696..891H, 2009ApJ...698.1550H, 2010ApJ...719.1315K}, and so the second treatment draws values of \fedd\ randomly at each step from a Schechter function \citep[power-law slope $\alpha=0.4$ and exponential cutoff at $0.4L_{\textrm{Edd}}$;][]{2009MNRAS.398..333H, 2016arXiv160501739J}.  The maximum allowed \fedd\ is unity (and the rapidly dropping distribution at high Eddington ratios means that values above unity would be extremely rare in any case), and the minimum cutoff is set by the desired \meanfedd.  At each time step the \fedd\ for a given object can change (i.e.\ the Schechter function is sampled at each step), allowing the quasar to ``flicker'' \citep[e.g.][]{2011MNRAS.416..650S, 2014ApJ...782....9H, 2015MNRAS.451.2517S}.  The mean \fedd\ is a free parameter (and in the more simple single \fedd\ case is simply the chosen value for every object).  We note that the mean \fedd\ we refer to is always the \emph{input} mean, while in the varying \fedd\ model after cuts are applied (see Section 2.2) the mock samples may have a very different mean Eddington ratio.  In all cases the BH growth phase continues until \mbh\ reaches its expected value, at which point the BH returns to dormant with a constant mass unless the mass ratio again falls outside \mswitch\ and another growth phase begins.  An illustration of the growth of a single source (including its halo, stellar, and black hole mass), is shown in Figure~\ref{fig:growth}.

An important assumption in our model is that the quasar phase begins with an obscured period, and later transitions to an unobscured phase.  This is consistent with some models of merger-driven AGN activity \citep[e.g.][]{1988ApJ...325...74S, 2005Natur.433..604D, 2005Natur.435..629S, 2008ApJS..175..356H, 2008MNRAS.386.2285C, 2010MNRAS.405L...1B}, and (as we argue below) provides a natural explanation for the observed higher halo masses of obscured sources.  While we require that the obscured phase comes first, we leave the relative fraction of time spent in the obscured phase (\tobsc\ $=t_{\textrm{obsc}}/t_{\textrm{AGN}}$) as a free parameter.  Figure~\ref{fig:growth} highlights where in the quasar phase the source is considered obscured (red) and unobscured (blue), for a given choice of \tobsc.  Note that the entire growth phase is not colored --- this is due both to the fluctuating \fedd\ as well as the luminosity cut described in the next section.

%In summary, at this point the model has three parameters that we can vary in order to explore their impact on the resultant samples: the ratio of the actual and expected black hole mass that determines when a BH enters and AGN phase (\mswitch), the lifetime of the obscured phase as a fraction of the total AGN growth period (\tobsc), and the mean Eddington fraction (\meanfedd), which is either a constant value or the mean of a random sampling from a Schechter function.  A forth and final critical parameter is introduced in the next subsection.

\subsection{Mock observed samples}
The relatively simple prescription above allows us to generate mock observed samples of quasars, to explore how the quasar population varies with the parameters \mswitch, \tobsc, and $\langle$\fedd$\rangle$.  Critically, we also explore the role of various luminosity cuts (\lcut), to mimic the behavior of certain method-dependent quasar selection effects and place our modeled results in the context of observed samples.  We run our growth simulations from $0 < z < 3$, but we restrict ``observed'' samples to the range $0.5 < z <  1.5$, where the bulk of objects selected from \textit{WISE} colors lie \citep[e.g.][D16a]{2012ApJ...753...30S, 2013ApJ...772...26A, 2014MNRAS.442.3443D, 2015MNRAS.446.3492D}. 

We generate simulated obscured and unobscured samples, which are simply defined based on the amount of time an object has currently been in an actively growing BH phase and the adopted value of $\tau_{\textrm{obsc}}$, with the earlier phases of growth defined as obscured. We use the following grid of parameters, each in steps of 0.1: $44 \leq \log L_{\textrm{cut}} \leq 47$, $0.1 \leq \Delta M_{\textrm{switch}} \leq 1$, $0.2 \leq \tau_{\textrm{obsc}} \leq 0.8$, and $0.1 \leq \langle \lambda_{\textrm{Edd}} \rangle \leq 1$.  For each parameter combination, we store the median halo, stellar, and black hole mass, median luminosity, and the maximum number of distinct SMBH growth phases an object goes through.

Observationally, the actual measured parameter (for example with clustering measurements) is not the median halo mass but the effective bias $b_{\textrm{eff}}$ and the corresponding effective  halo mass $M_{\textrm{h,eff}}$.  The median and effective halo mass are not strictly equal, due to the strong mass dependence of the bias.  Therefore, we also calculate $M_{\textrm{h,eff}}$ for each parameter set in the following way.  A cubic spline is fit to the normalized halo mass distribution in order to approximate $dN/dM$.  This can be combined with $b(M)$ to determine the effective bias at $z=1$, the mean redshift range of interest: 
\begin{equation}
b = \frac{\int b(M) \frac{dN}{dM} dM}{\int \frac{dN}{dM} dM}.
\end{equation}
Using the relationship between bias and halo mass from simulations \citep[][]{2010ApJ...724..878T}, and the matter power spectrum generated with \textsc{camb} \citep{2000ApJ...538..473L}, the effective bias is converted into the effective mass at $z=1$ (see Section 3.5 of D16a and the associated code libraries). Because of the shape of the halo mass distributions (see Section 3), this effective mass tends to be larger than the median halo mass by on average 0.1 to 0.2 dex.

\section{Results \& Discussion}
For simplicity, we first explore the properties of the mock observed samples using the more simple constant \fedd\ model, where a single Eddington ratio is assigned to all objects at all times.  We will discuss the impact of the varying \fedd\ on these results in Section 3.4.

\subsection{Halo mass versus \lcut, \mswitch, \tobsc, and \meanfedd}
\begin{figure*}
\centering
   \includegraphics[width=15cm]{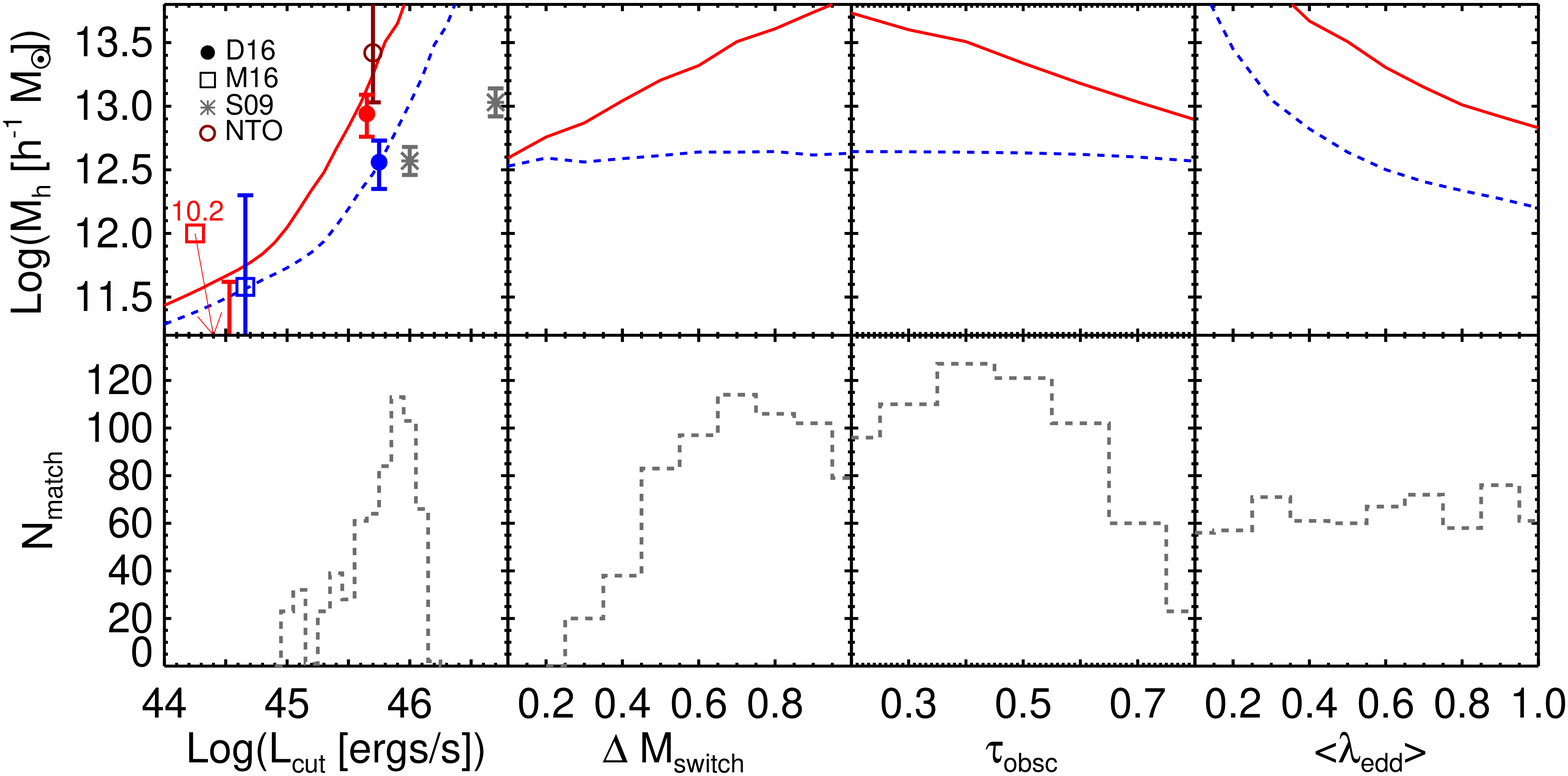}
  \caption{Model results when the Eddington ratio is a single value for all objects at all steps, equal to the chosen \meanfedd.  \emph{Top row:} The resulting obscured (solid red) and unobscured (dashed blue) effective dark matter halo mass, as a function of the luminosity cut, the actual to expected mass ratio at which BH growth begins (\mswitch), the obscured lifetime, and the mean Eddington ratio.  In each panel, the other three parameters are held constant at $L_{\textrm{cut}} = 45.8$, $\Delta M_{\textrm{switch}}=0.7$, $\tau_{\textrm{obsc}}=0.4$, and $\langle \lambda_{\textrm{Edd}} \rangle = 0.5$.  The first panel shows some recent measurements for comparison with the model --- see Section 3.2 for details.  ``D16'' represents the observed obscured and unobscured halo masses from D16a, ``M16'' are the obscured and unobscured real-space measurements of \citet[][note that the obscured halo mass is below the scale of the $y$-axis, though the top of the error bar is visible, and its value is indicated above the red square with an arrow]{2016ApJ...821...55M}, ``S09'' are the luminosity-dependent unobscured measurements of \citet{2009ApJ...697.1656S}, and ``NTO'' is the ``non-torus obscured'' modeled subset of the obscured population from D16b.  \emph{Bottom:} The distributions of parameters that provide  matches within the errors (see section 3.2) to the results of D16a and D16b.\label{fig:cuts}}
\end{figure*}

In the top row of Figure~\ref{fig:cuts}, we show the effective halo mass of the unobscured (blue) and obscured (red) samples as a function of our four free parameters.  When one parameter is varied, we hold the other values constant at $L_{\textrm{cut}} = 10^{45.8}$ ergs s$^{-1}$, $\Delta M_{\textrm{switch}} = 0.7$, $\tau_{\textrm{obsc}} = 0.4$, and $\langle \lambda_{\textrm{Edd}} \rangle = 0.5$.  These are chosen to highlight agreement with observations (see Section 3.2), but also represent reasonable values for each of the parameters.  The bolometric luminosity of mid-IR selected samples is quite high, generally peaking above $10^{46}$ ergs s$^{-1}$ \citep[e.g.][]{2007ApJ...671.1365H, 2013ApJ...772...26A}.  Highly luminous \emph{WISE}-selected quasars also tend to radiate at a large range of Eddington ratios, so a mean of 0.5 is not unlikely \citep[e.g.][]{2013ApJ...772...26A}.  D16a estimated that the lifetime of the obscured phase is roughly equal to the unobscured lifetime (though the uncertainty is large), which is generally consistent with models \citep[e.g.][]{2008ApJS..175..356H}.  The chosen value of \mswitch\ leads to growth periods that last on the order of 200 Myr, with each source typically going through two growth phases from $0 < z < 3$. 

\begin{figure}
\centering
   \includegraphics[width=8cm]{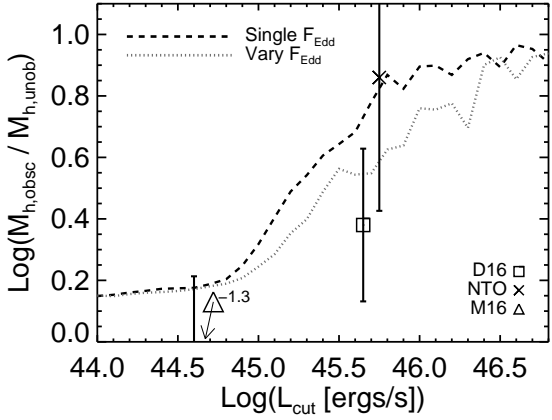}
  \caption{The ratio of the obscured and unobscured effective halo masses, as a function of \lcut, for the fixed \fedd\ case (dashed black line) and with a varying \fedd\ (dotted grey line).  While the obscured sample always has a higher halo mass at any \lcut, the difference is very small at low luminosity and increases rapidly above $10^{45}$ ergs s$^{-1}$, reaching nearly a full dex at the highest luminosity.  Overlaid are mass ratios from observed obscured and unobscured samples from D16a, the modeled ``non-torus obscured'' sample of D16b, and \citet{2016ApJ...821...55M}. Note that \citet{2016ApJ...821...55M} found that the obscured halo masses were \emph{lower} than those of the unobscured population (so the mass ratio is negative here), though with the large errors they were consistent with being the same, and are consistent with our model. \label{fig:massratio}}
\end{figure}

The tracks in the first panel of Figure~\ref{fig:cuts} illustrate first the qualitative ability of this simple model to separate the halo masses of the two samples.  It is clear that the halo masses of the obscured sample are larger than those of the unobscured sample (and we point out the shape of this relationship mimics the \mhalo\ - \mbh\ relationship, since the luminosity and black hole mass are directly related).  This is true at any \lcut\ probed here, but the difference in effective halo mass increases as the luminosity cut increases.  This is seen more directly in Figure~\ref{fig:massratio}, which shows the mass ratio between obscured and unobscured samples as a function of \lcut\ (the black dashed line is for the single \fedd\ case considered here), and illustrates that the mass difference can reach a full dex at high luminosity.  Since the luminosity is directly tied to the black hole mass (via \fedd), and the unobscured phase is always the later portion of the growth period, the luminosity for a given object is always higher in this phase because the black hole mass is higher.  Therefore, at a given \lcut, the obscured sources must be found, on average, in higher mass halos because the objects in lower mass halos are missed.  There are unobscured sources in high mass halos as well, but they can be detected in the far more common lower mass halos, shifting their distribution toward lower masses.  This effect separates the effective masses of the populations.

In the three right-most panels in the top row of Figure~\ref{fig:cuts}, we illustrate the behavior of the halo mass with other parameters of the model.  The unobscured halo mass is only very weakly dependent on \mswitch\ and \tobsc, while the obscured halo mass increases with the former and decreases with the latter.  This is simply because the luminosity cut behaves effectively as a black hole mass cut, and again the obscured phase occurs (for a given object) when its black hole mass is lower.  By making \mswitch\ larger, more of the obscured phases happen at lower BH mass, which are not observed due to the luminosity cut, and the obscured haloes become biased toward larger masses.  The inverse happens as \tobsc\ increases, because it increases the range in black hole mass that the obscured phase reaches, thus allowing more sources at lower halo mass to be selected in the obscured phase. Finally, as \meanfedd\ increases, the halo masses of both samples decrease.  This is because at higher \fedd, lower-mass black holes, and therefore lower-mass dark matter haloes, will satisfy the luminosity cut.

\subsection{Comparison with observations}
In the first panel of Figure~\ref{fig:cuts} and in Figure~\ref{fig:massratio} we include some recent halo mass measurements, summarized below and in Table~\ref{tbl:observations}, with which to compare our model quantitatively.  These include the \citet{2009ApJ...697.1656S} optically selected quasars split by the 10\% most and 90\% least luminous sources, the D16a IR-selected obscured and unobscured CMB lensing-cross correlation measurements, the D16b non-torus obscured modeled result, and the \citet{2016ApJ...821...55M} real-space clustering measurements (adopting their ``Assef IR AGN'' measurements that exclude the COSMOS field), which are selected in a similar way as in D16a.  We point out that \citet{2016ApJ...821...55M} actually find that unobscured quasars reside in marginally higher mass haloes, though the results are consistent with there being no difference.

In all cases bias values from the relevant reference are converted to halo masses using the \citet{2010ApJ...724..878T} form for $b(M)$, as well as our cosmology, matter power spectrum, and code\footnote{See \url{https://github.com/mdipompe/halomasses}} (D16a) for consistency.  For convenience, we list the biases and our calculated halo masses for these data in Table~\ref{tbl:observations}.  Note that the low halo masses for the \citet{2016ApJ...821...55M} samples, which, combined with the smaller sample sizes, serve to increase the error bars significantly.  This is largely due to the fact that at lower halo masses, $b(M)$ becomes quite flat and a given bias can be consistent with a much larger range of halo mass.  These large error bars reduce the predictive power of our model somewhat, and more detailed studies at lower luminosities can help better constrain its predictions.  While the mean redshifts of the samples vary, they all include quasars at $z \sim 1$.

\begin{table}
\centering
  \caption{Bias and halo masses from the literature}
  \label{tbl:observations}
  \begin{tabular}{lcccc}
  \hline
        Sample           &  &   $\langle z \rangle$ & $b_q$             & $\log(M_h/M_{\odot} \ h^{-1})$   \\
  \hline
  D16 unobscured   &  & 1.05 &1.72 $\pm$ 0.18  & $12.56_{-0.21}^{+0.17}$    \\
  D16 obscured       &  & 0.98 & 2.06 $\pm$ 0.22  & $12.94_{-0.18}^{+0.15}$     \\
  NTO                     &  & 1.0 & 3.08  $\pm$ 0.93 &  $13.42_{-0.39}^{+0.39}$    \\
  M16 unobscured  &  & 0.70 & 0.96 $\pm$ 0.27 &  $11.58_{-2.01}^{+0.72}$    \\
  M16 obscured      &  & 0.77 & 0.75 $\pm$ 0.25 &  $10.21_{-4.21}^{+1.41}$    \\
  S09 highest $L$ (unob)  &  & 1.4 & 3.00 $\pm$ 0.25 &  $13.03_{-0.11}^{+0.11}$    \\
  S09 lower $L$ (obsc)     &  & 1.4 & 2.22 $\pm$ 0.14 &  $12.57_{-0.11}^{+0.11}$    \\
\hline
   \end{tabular}
   \\  
{
\raggedright    
 Bias values and inferred halo masses used for comparison with the model in Figures~\ref{fig:cuts},~\ref{fig:massratio}, and~\ref{fig:cuts2}, along with the mean redshift of each sample. The biases are taken directly from the indicated references --- D16 indicates \citet{2016MNRAS.456..924D}, NTO indicates the simulated ``non-torus obscured'' samples of \citet{2016MNRAS.460..175D}, M16 indicates \citet{2016ApJ...821...55M}, and S09 indicates \citet{2009ApJ...697.1656S}. For consistency in comparisons, all halo masses are calculated using our cosmological parameters and procedures \citep{2016MNRAS.456..924D}.  Note that lower masses tend to have larger errors for the same bias error, due to the shape of the $b(M)$ relationship.  This leads to very large errors, especially on the lower end, for the \citet{2016ApJ...821...55M} halo masses. \\
 }
\end{table}

Note that most of these works do not present the minimum luminosities of their samples, but rather the means or medians, and so their position along the \lcut\ (the lower limit of the luminosities) axis is approximate, based on distributions shown in the reference of interest and are likely accurate to of order $\sim$0.1 dex.  Bolometric luminosities for the \citet{2009ApJ...697.1656S} data are adopted directly from their work, while for the \citet{2016ApJ...821...55M} data we apply a constant bolometric correction of 12 to their IR luminosities for an approximate $L_{\textrm{bol}}$ \citep[e.g. Figure 12 of][]{Richards:2006p3932, 2007ApJ...671.1365H}.  Bolometric luminosities for the D16a samples are based on SED fits from sources selected in a similar way in the B\"{o}otes field \citep{2011ApJ...731..117H}.  Where bolometric corrections are applied additional scatter is present that adds uncertainty to the minimum luminosity estimates. Typical bolometric corrections in the IR have scatter of order a few \citep[e.g.][]{2012MNRAS.426.2677R}, which propagates to a scatter in luminosity of a few tenths of a dex.

In addition to the qualitative behavior of the halo masses, this simple model agrees fairly well with real data using reasonable parameter choices.  There is excellent agreement with D15a, particularly the unobscured sample, and for this parameter set the agreement with the NTO halo masses of D16b is excellent.  Figure~\ref{fig:massratio} highlights the better agreement in the obscured to unobscured mass ratio when the NTO obscured, rather than the full observed obscured, sample is used.  We also see in Figures~\ref{fig:cuts} and~\ref{fig:massratio} that the model agrees with the measurements of \citet{2016ApJ...821...55M}.  This is simply due to the different luminosity regimes probed by the sample --- because \citet{2016ApJ...821...55M} probe down to lower luminosities, it is expected that they should find a far weaker dependence on halo mass with quasar type.  At these low luminosities, the difference in halo mass is only $\sim$0.1 dex, far smaller than typical measurement uncertainties.  Therefore, our model shows that these two measurements may not really be in disagreement.  

\citet{2009ApJ...697.1656S} find no strong evidence for a dependence of clustering strength (and thus halo mass) on bolometric luminosity, except when considering only the most luminous 10\% of objects.  Our model agrees well with the measurement for the 90\% of objects below these highest luminosities, but over-predicts the masses for the most luminous sample.   Our model also predicts a dependence on luminosity over the full dynamic range, which does weaken toward lower luminosities, which was not seen by \citet{2009ApJ...697.1656S}.  We note that our model has not included scatter in the \mhalo\ - \mstar\ - \mbh\ relationships --- both for simplicity and because quantifying the intrinsic scatter is difficult.  At high luminosity and high mass, well above the ``knee'' in both the halo mass function and quasar luminosity function, this scatter is much more likely to shift low mass objects to higher masses or luminosities rather than vice versa, flattening the \mhalo\ - \lcut\ relationship.  As a check we added Gaussian scatter with a standard deviation of 0.2 dex to the expected BH masses, and find that it does flatten the relationship slightly (while adding several other complicating factors), though not enough to match the high luminosity \citet{2009ApJ...697.1656S} data point.

In order to identify regions of our parameter space that can reproduce observations, we search our results for outputs with effective halo masses between $13.0 < M_{\textrm{h,obsc}} < 13.8$ $M_{\odot}/h$ (the predicted NTO effective halo mass; D16b) and $12.35 < M_{\textrm{h,unob}} < 12.73$ (the observed unobscured effective halo mass; D16a).  In the bottom panels of Figure~\ref{fig:cuts}, we show histograms of the parameters that produce such a match.  We note that these parameters are highly dependent on one another, but point out that there are many combinations that can produce broad agreement with observations.  Notably, as long as the other parameters are properly constrained, each variable can generally have a wide range of values and still agree with observations.  \lcut\ generally falls between $\sim$10$^{45 - 46}$ ergs s$^{-1}$, in agreement with the observed low end of the \emph{WISE}-selected quasar luminosity distribution.  The obscured lifetime and mean intrinsic Eddington ratio can span the full range of values, and the value of \mswitch\ generally needs to fall above $\sim$0.3.  We note that for $0.3 \leq \Delta M_{\textrm{switch}} \leq 0.7$ most sources go through, on average, two major growth phases over $0 < z < 3$, while for higher \mswitch\ only one growth period occurs.  This may provide additional predictions for the value of \mswitch, though in practice determining the number of distinct growth phases is quite difficult, especially in the presence of flickering within a growth period even if the total quasar lifetime is known \citep[e.g.][]{2008MNRAS.391.1457K, 2011ApJ...736...49L, 2012MNRAS.424..933W}.  

\begin{figure*}
\centering
   \includegraphics[width=12cm]{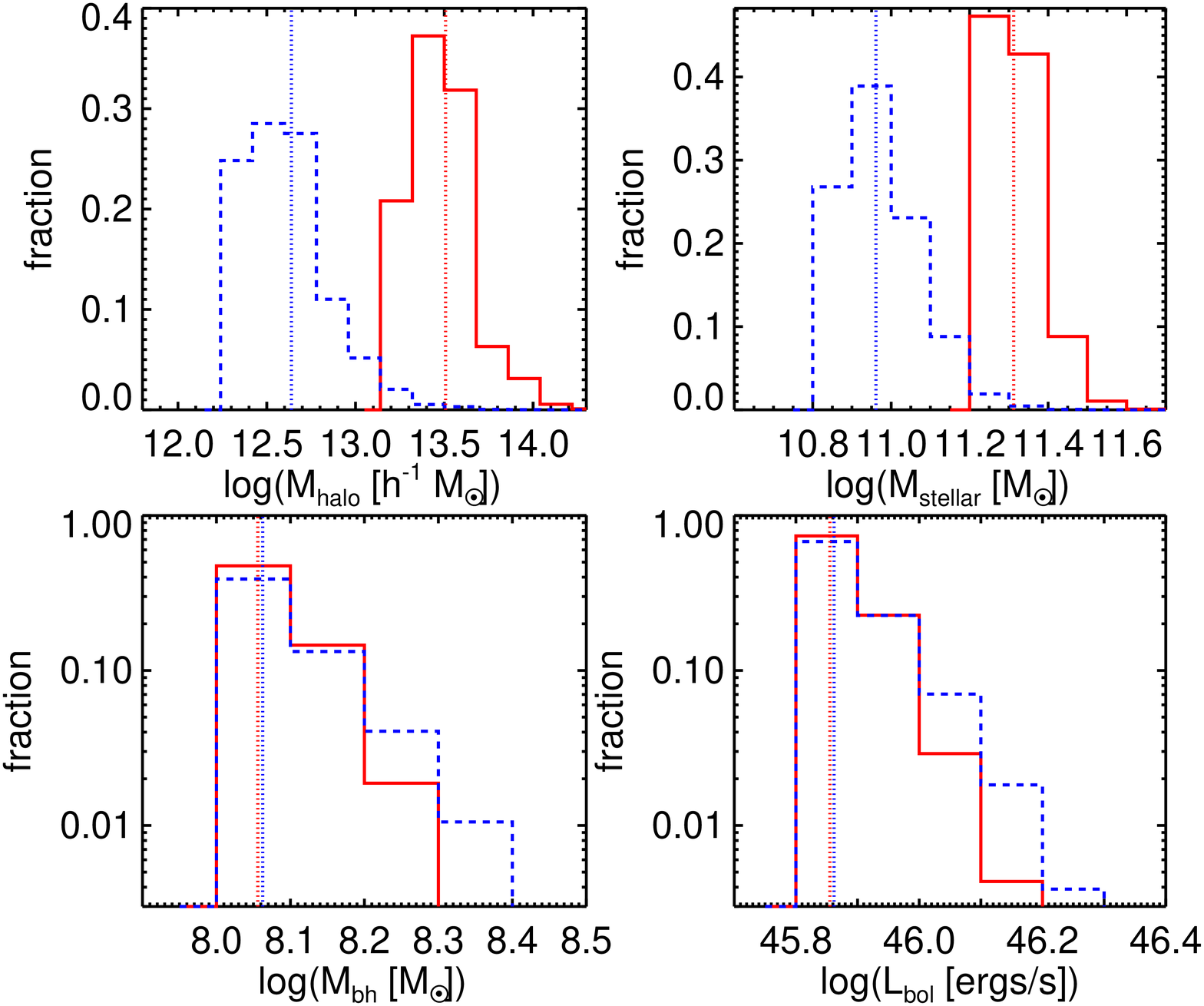}
  \caption{Output mass and luminosity distributions for the fixed \fedd\ model and $\log (L_{\textrm{cut}})=45.8$ ergs s$^{-1}$, $\Delta M_{\textrm{switch}}=0.7$, $\tau_{\textrm{obsc}}=0.4$, and $\langle \lambda_{\textrm{Edd}} \rangle = 0.5$, as in Figures~\ref{fig:cuts} and~\ref{fig:massratio}. Note the log $y$-axes in the bottom panels.  The luminosity cut (bottom right) results in a clear separation of the halo (top left) and stellar (top right) mass distributions, while keeping similar black hole (bottom left) mass distributions.  Dashed lines in the \mhalo\ panel indicate the effective mass inferred from a bias measurement of these mass distributions, and median values in all other panels.\label{fig:out_masses}}
\end{figure*}

In Figure~\ref{fig:out_masses}, we show the observed distributions of masses (\mhalo, \mstar, and \mbh) and $L_{\textrm{bol}}$ for the fixed parameters listed above ($L_{\textrm{cut}} = 10^{45.8}$ ergs s$^{-1}$, $\Delta M_{\textrm{switch}} = 0.7$, $\tau_{\textrm{obsc}} = 0.4$, and $\langle \lambda_{\textrm{Edd}} \rangle = 0.5$).  In the \mhalo\ panel, the dashed lines mark the effective masses, while in the other panels dashed lines mark the median values.  We see that the halo mass distribution for the obscured sample is offset to higher masses, but generally has a similar shape as the unobscured sample.  The shape of these distributions is roughly consistent with predictions from halo occupation distribution (HOD) analyses of unobscured quasars, which find that quasar hosts have a log-normal mass distribution \citep[e.g.][]{2012ApJ...755...30R, 2013ApJ...779..147C}.  
%Similar analysis has not yet been performed for obscured quasars, but \citet{2016MNRAS.460..175D} utilized mass distributions to estimate the non-torus obscured typical halo mass that are quite similar to those generated from the model here.  
Figure~\ref{fig:out_masses} also shows that this model predicts that the stellar masses of the obscured population will be higher, and that this is a driver for the difference in halo mass.  However, the resulting black hole masses are around typical values of $10^8$ M$_{\odot}$, and quite well matched between the two populations.  This results in well-matched luminosities as well, given the similar Eddington ratios.

\subsection{The obscured fraction}
\begin{figure*}
\centering
   \includegraphics[width=15cm]{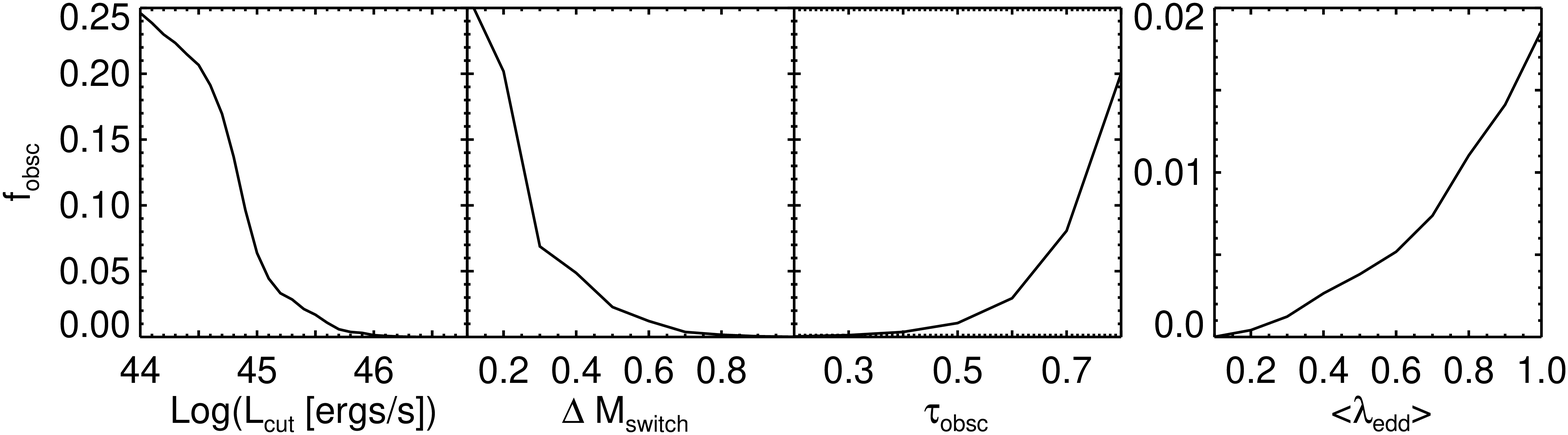}
  \caption{The obscured fraction as a function of \lcut, \mswitch, \tobsc, and \meanfedd.  When one parameter is varied, others are held constant with $\log (L_{\textrm{cut}})=45.8$ ergs s$^{-1}$, $\Delta M_{\textrm{switch}}=0.7$, $\tau_{\textrm{obsc}}=0.4$, and $\langle \lambda_{\textrm{Edd}} \rangle = 0.5$, as in Figures~\ref{fig:cuts} and~\ref{fig:massratio}.  When using parameters that produce a large separation in obscured and unobscured halo mass, as predicted for the non-torus obscured sample in D16b, the obscured fraction drops below expected values.  The value of \meanfedd\ does not strongly affect \fobsc, and so varying it while keeping other parameters fixed results in a consistently low value --- note the smaller range in the $y$-axis of the \meanfedd\ panel.  However, small tweaks to these four parameters can increase the obscured fraction to more realistic values, while slightly reducing the predicted difference in halo masses.\label{fig:fobsc}}
\end{figure*}

The model not only predicts halo mass as a function of the various parameters, but also an observed obscured quasar fraction (\fobsc).  We show \fobsc\ as a function of each of our parameters (holding the same fixed values in each panel as above) in Figure~\ref{fig:fobsc}.  Broadly speaking, the value of \fobsc\ behaves in the opposite manner as \mhalo\ with each variable, simply because more massive haloes are more rare.  As discussed in D16b, the primary driver of the difference in halo mass is not the complete observed obscured population, but the NTO subset that is intrinsically different from the unobscured population.  The relative fraction of this population is difficult to constrain with current observations, but D16b argue that the best current estimate is $\sim$10\% of the full population (25\% of the obscured population).  While the model is able to produce matches to the halo masses with the combination of parameters used in the previous figures, the obscured fractions tend to be biased very low, on the order of $\sim$1\%.  

Adjusting the parameters slightly to $\log$\lcut$=$45.7 ergs s$^{-1}$, \mswitch$=$0.5, \tobsc$=$0.6, and \meanfedd$=$0.3, all reasonably within expected ranges, produces an obscured fraction of 8\%.  This does however shift the unobscured halo masses upwards by $\sim$0.1 dex, at the high end but within the error bar of D16a. It also lowers the obscured masses to within the upper error bar of the observed obscured halo masses in D16a and the lower error bar of the NTO sample of D16b.  The obscured to unobscured mass ratio is then $\sim$2.5.  Therefore there is a slight tension in our model between matching the observed effective halo masses and matching the obscured fraction.  In order to raise the obscured fraction, the difference between the halo masses must necessarily decrease.  However, even with the simplicity of the model, we are able to find broad agreement with observations.

\subsection{Results using a distribution in \fedd}
\begin{figure*}
\centering
   \includegraphics[width=15cm]{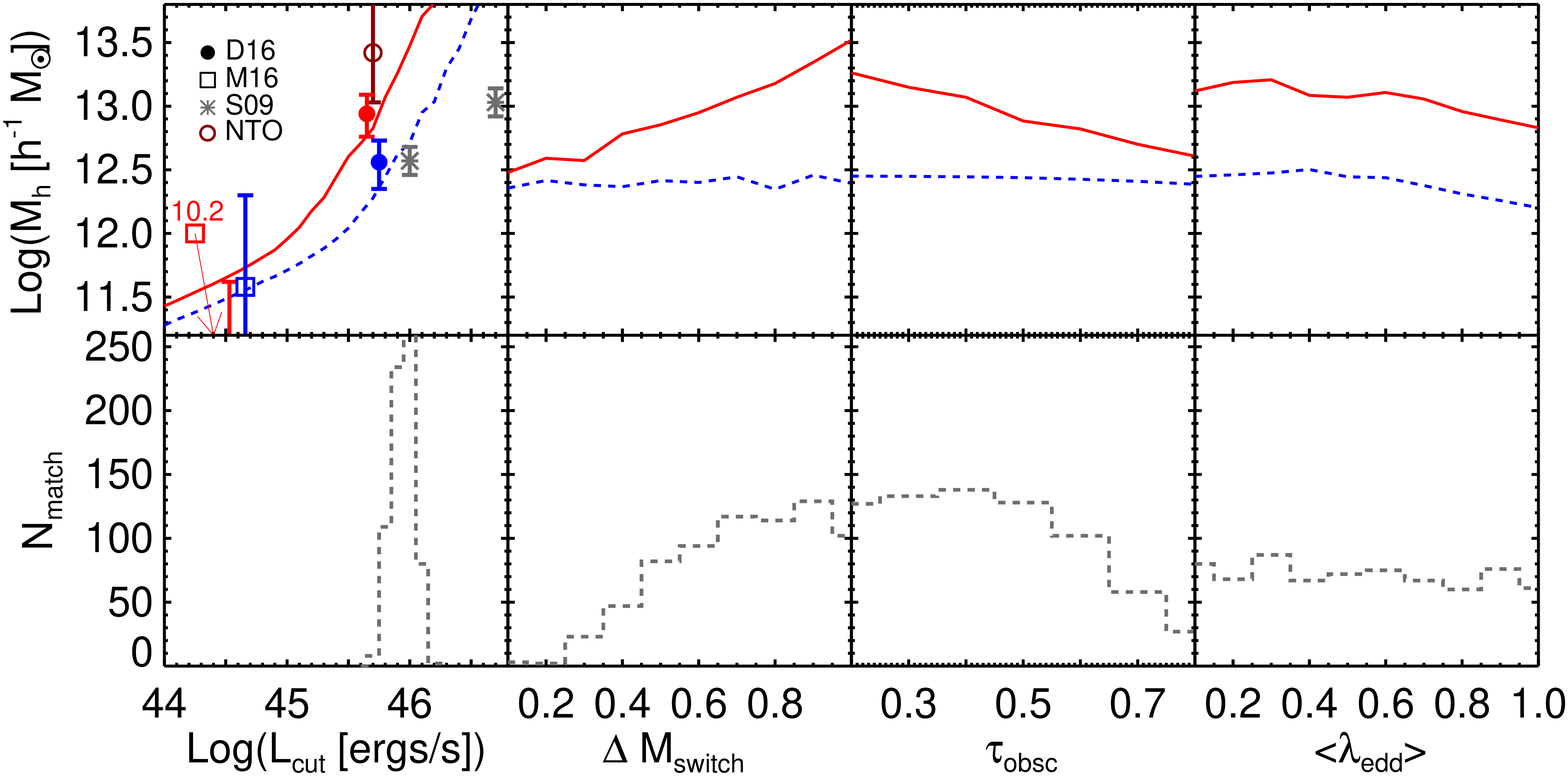}
  \caption{The same as Figure~\ref{fig:cuts}, but using a model with a distribution of \fedd\ values, rather than adopting a single mean value for each object at each time.  In the right-most panels, ``Mean \fedd'' refers to the mean of the intrinsic (input) distribution, not necessarily the mean \fedd\ of the resulting mock sample.  In fact, the mean of the observed sample is always quite high (above $\sim$0.8), because more common low-mass black holes (and haloes) are scattered above the \lcut\ by being assigned a high \fedd, and these dominate the distribution.  This explains the lack of strong dependence on the input \meanfedd\, as well as the lower halo masses in the other panels for the same parameter combinations in Figure~\ref{fig:cuts}.\label{fig:cuts2}}
\end{figure*}

We now turn to the effects of including an Eddington ratio distribution, as opposed to assuming a singular mean value for the entire population.  We show these results in Figure~\ref{fig:cuts2}, which has the same panels and values for fixed parameters as Figure~\ref{fig:cuts}, and with the grey dotted line in Figure~\ref{fig:massratio}.  The overall behavior with most parameters is the same --- raising \lcut\ increasingly separates the halo masses of the obscured and unobscured populations (though the increase in the separation is less rapid), increasing \mswitch\ enhances the halo mass difference, and increasing \tobsc\ has the opposite effect.  

However, the dependence on \meanfedd\ is far weaker.  The reason for this is that the high \fedd\ portion of the distribution allows lower mass black holes to satisfy the luminosity cut.  Because these objects are more numerous, they begin to dominate the distribution.  While the \emph{intrinsic }\fedd\ distribution is changing with \meanfedd\, the actual \emph{selected} population is dominated by those that have high \fedd.  Regardless of the input \meanfedd\, the resulting sample does not change significantly, causing a much flatter \mhalo\ - \meanfedd\ relationship.

With lower mass black holes and dark matter haloes playing a more significant role, it is more difficult to raise the halo masses of the unobscured population by changing \meanfedd\ (the parameter that has the largest effect on \mhalo\ for the unobscured sources, see Figure~\ref{fig:cuts}). However, Figure~\ref{fig:cuts2} shows that this model is within the lower range of measured unobscured halo masses.  For the parameter combination shown in Figure~\ref{fig:cuts2} the obscured halo mass trend intersects the error bar of the observed obscured value and is farther from the predicted NTO value.  By increasing \mswitch\ and/or decreasing \tobsc\ the model can match the NTO halo masses fairly well. However, Figure~\ref{fig:massratio} shows that, given the combined errors, this model including an Eddington ratio distribution is consistent with all of the observed mass ratios.  We also see in the bottom panels of Figure~\ref{fig:cuts2} that there are several parameter ranges than can roughly match observations, but in general \lcut\ needs to be slightly higher in this version of the model to do so.

Finally, it is possible to match the obscured fraction of $\sim$10\% with the \fedd\ distribution model.  However, this generally requires a luminosity cut above $10^{46}$ ergs s$^{-1}$.  Though this is where the $L_{\textrm{bol}}$ distribution for \emph{WISE}-selected quasars peaks, there is likely a significant fraction below this value.  Therefore, like the single \fedd\ model, there is tension here between the predicted halo masses and obscured fraction, though it is possible to match them in a broad sense.

\section{Summary \& Conclusions}
We have shown that a simple model for the growth of dark matter haloes, galaxies, and supermassive black holes, combined with an evolutionary quasar sequence from obscured to unobscured, can both qualitatively and quantitatively reproduce recent observed differences in the obscured halo mass.  By incorporating a luminosity cut, which is present in flux-limited samples such as those derived from \emph{WISE}, we provide a natural explanation for seemingly disparate results that probe slightly different luminosity regimes.  The primary tension in the model appears to be in predicting both the effective halo masses of the obscured and unobscured population alongside the obscured fraction of the population.  However, broad agreement with observed values of both parameters is possible, even with this simple model.  

As we more accurately probe the halo masses of the obscured population with improved and ever larger samples, incorporating redshift information and analyses of the full halo occupation distribution, this simple model provides a compelling framework for interpreting the results.  Relying on only a few empirical relationships, we have illustrated that an evolutionary component to the obscured quasar phenomenon may be necessary to explain all observations, in addition to the role of orientation and line-of-sight effects.

\section*{Acknowledgements}
MAD, RCH, and ADM were partially supported by NASA through ADAP award NNX12AE38G.  MAD and RCH were also partially supported by the National Science Foundation through grant numbers 1211096 and 1515364.  MAD and ADM were also supported by NSF grant numbers 1211112 and 1515404.  RCH also acknowledges support from an Alfred P. Sloan Research Fellowship, and a Dartmouth Class of 1962 Faculty Fellowship.  JEG is supported by the Royal Society.

\bibliography{/Users/Mike/Dropbox/full_library}

\label{lastpage}

\clearpage

\end{document}